%Paper: hep-th/9302087
%From: <GRIGORE%ROIFA.BITNET@pucc.Princeton.EDU>
%Date: Fri, 19 Feb 1993 10:19 +0200

\magnification=1200
\noindent
\pageno=1
\centerline{ON THE UNIQUENESS OF THE NEWTON-WIGNER POSITION OPERATOR}
\centerline{D.R. Grigore}
\centerline{Department of Theoretical Physics, Institute of
Atomic Physics}
\centerline{Bucharest-Magurele,Romania}

{\bf 1. Introduction}

The notion of localizability of relativistic particles has
attracted considerable attention. It is generally admitted
that the proper mathematical setting is the one due to
Wightman [1] (see also [2]) which originates from the physical
ideas of Newton and Wigner.

Let us out line briefly this framework. We admit that the
configuration space of a certain physical system is the Borel
space $(Q,\beta)$; here $\beta$ is the Borel structure
on $Q$. We also consider that the system is a pure quantum
one. Then the lattice of the system is of the form
$P({\sl H})$; here ${\sl H}$ is some Hilbert space and $P({\sl H})$ is
the lattice of orthogonal projectors in ${\sl H}$. (see for
instance [2]).

Then one can argue (see [1] pg. 847) that the position
observable is a projection-valued measure:
$$\beta \ni E \rightarrow P_{E} \in P({\sl H})$$

The physical interpretation is the following: the
states in the range of $P_{E}$ correspond to
localization of our system in $E \subseteq Q$.

If $G$ is a (Borel) group of symmetries, the there exists
in $H$ a projective unitary representation of $G$:
$$G \ni g
\rightarrow U_{g} \in U({\sl H})$$

Here $U(H)$ is the set of unitary operators in ${\sl H}$.

If $Q$ is a $G$-space, then a natural compatibility
condition is:
$$U_{g} P_{E} U_{g}^{-1} = P_{g\cdot E}.\eqno(1.1)$$

Here $g\cdot E$ is the image of $E$ under the action
of $G$. The couple $(U,P)$ is called a system of
imprimitivity for $G$ based on $Q$.

If $P$ is a position observable on $Q = R^{3}$,
then according to the spectral theorem, we can
construct three self-adjoint commuting operators
$Q_{1},Q_{2},Q_{3}$ namely the position operators
appearing in the Newton-Wigner analysis. A convenient
way to construct them [2] is to define first the
unitary operators:
$$B({\bf y}) \equiv \int exp(-i{\bf x} \cdot {\bf y})
dP({\bf x}).\eqno(1.2)$$
for any ${\bf y} \in R^{3}$. Then, $Q_{k}$ ($k =1,2,3$)
are determined
via Stone theorem by:
$$Q_{k} f \equiv i {d \over dt} B(t{\bf e}_{k})f
\vert_{t=0}.\eqno(1.3)$$

Here ${\bf e}_{k}$ ($k=1,2,3$) is the canonical basis
in $R^{3}$.

A very interesting situation appears when one tries
to combine these ideas with relativistic invariance.
Suppose that our system admits a group of invariance
$G$ which has as a subgroup the special Euclidean
group $SE(3)$. Then, we must have a projective unitary
representation of $G$ in ${\sl H}$. By restriction, we have
a projective unitary representation $U$ of $SE(3)$
in $H$. One says that the system is localisable
(in $R^{3}$) {\it iff} there exists a position
observable $P$ (based on $R^{3}$) such that
$(U,P)$ is a system of imprimitivity.

According to Newton-Wigner-Wightman analysis
the non-zero mass systems are localizable and
zero mass systems are not localizable in this sense.
The non-zero and zero mass systems are by definition
certain irreducible unitary continous representations
of the Poincar\'e (or the Galilei) group. The non-localizability
for zero mass systems, especially for the photon, has
generated a rather extensive literature [4]-[10].

On the other hand, even in the case of non-zero
mass systems (which are localizable) a curious
phenomenon appears, namely the position observable
is not unique. In some cases this arbitrariness
can be explicitely described [2]. Namely, suppose
that $P^{0}$ is a position observable of finite
multiplicity and $P$ is an arbitrary position
observable. Then, there exists an unitary operator
$A$ in ${\sl H}$ such that $V$ commutes with the representation
$U$ of $SE(3)$ and:
$$P = AP^{0}A^{-1}.\eqno(1.4)$$

This non-unicity problem seems to be considered not
very bothersome. In [2] one can fine a side remark
to the sense that this problem is ``noteworthy''
and there exists a particular choise of $P$ which
seems to be the ``simplest'' and which gives the classical
relationship between velocity and momentum.

On the other hand, it is known that the similar unicity problem
can be solved in the framework of classical Hamiltonian
mechanics, imposing the so-called ``manifest covariance''
condition [11]. Namely, if $K_{i}$ ($i = 1,2,3$) are
the (Lorentz) boosts generators and $H$ is the Hamiltonian,
then one can show that $Q_{1},Q_{2},Q_{3}$ behaves as the
spatial components of a quadrivector {\it iff} one has
the relation:
$$\{K_{i},Q_{j}\} = Q_{i}\{H,Q_{j}\}.\eqno(1.2)$$
for $i = 1,2,3$. It is tempting to use something
similar in quantum mechanics. In [12], one applies
the naive correspondence rules:
$$\{\hskip 0.5truecm ,\hskip 0.5truecm \} \rightarrow
i[\hskip 0.5truecm ,\hskip 0.5truecm ].\eqno(1.3)$$
$$AB \rightarrow 1/2 (\hat{A}\hat{B} + \hat{B}\hat{A}).\eqno(1.4)$$
and obtains for the quantum operators $K_{i}$ ,$H$ and
$Q_{i}$ the relation:
$$[K_{i},Q_{j}] = 1/2 \left(Q_{i}[H,Q_{j}] + [H,Q_{j}]
Q_{i} \right).\eqno(1.5)$$

Surprisingly, this equation gives an unique solution
for systems of zero-spin and has no solution for systems
of non-zero spin.

In this note, we propose a different solution. Namely,
we will impose a ``manifest covariance'' condition by a
purely quantum argument, without invoking the correspondence
rules (1.3) and (1.4). We will find an unique position
operator for any spin, which is exactly the one considered
in [2] as the ``simplest'' one. Moreover, we will show that
the relation (1.5) is valid up to terms of order $\hbar c^{-2}$.
This quantum correction is zero {\it iff} the spin is zero.
This completely explains the result of [12]. Moreover, this
indicates that for the Galilei group, this quantum correction
must dissapear.

We mention that these exists a number of attempts in the literature
to define a Lorentz covariant quadrivector position operator
(see e.g. [13]) but they seem to be different from ours as motivation
and outcome.

In Section 2 and 3 we analyse the case of Poincar\'e
and respectively Galilei invariance and in Section 4 we formulate
some conclusions.
\vskip 1truecm
{\bf 2. The Position Operator for Non-zero Mass Poincar\'e Systems}

2.1 In the notations of [2], the system $[m,s]$
of mass $m$ and spin $s$ corresponds to the irreducible
projective representation $W^{m,+,s}$ of the proper
orthochronous Poincar\'e group $P^{\uparrow}_{+}$.

We can realize this representation in the Hilbert space
${\sl H} = {\sl L}^{2}\left(R^{3},C^{2j+1},{d{\bf p} \over
E({\bf p})}\right)$ where:
$$E({\bf p}) \equiv \sqrt{{\bf p}^{2} + m^{2}}.\eqno(2.1)$$
as follows. Let $X^{+}_{m}$ be the positive energy
hyperboloid:
$$X^{+}_{m} \equiv \{p \in R^{4} \vert p_{0}^{2} -
{\bf p}^{2} = m^{2}, p_{0} > 0\}.\eqno(2.2)$$

We define $\tau:R^{3} \rightarrow X^{+}_{m}$ by:
$$\tau ({\bf p}) \equiv (E({\bf p}),{\bf p}).\eqno(2.3)$$
and $\eta:X^{+}_{m} \rightarrow R^{3}$ by:
$$\eta (p) \equiv {\bf p}.\eqno(2.4)$$

Then we have for any $a \in R^{4}$ and any $A \in SL(2,C)$
$$(W^{m,+,0}_{A,a} f)({\bf p}) = e^{ia \cdot \eta({\bf p})}
D^{(s)}(U(A,{\bf p}))
f(\eta (\delta(L)^{-1}(\tau({\bf p})))) .\eqno(2.5)$$
for any $f \in {\sl H}$. Here $U(A,{\bf p}) \in
SU(2)$ is the so-called Wigner rotation,  $D^{(s)}$
is the representation of weight $s$ of $SU(2)$ and
$\delta :SL(2,C) \rightarrow L^{\uparrow}_{+}$ is the
covering homomorphsim.

2.2 As we have said in the Introduction, the system
$[m,s]$ is localizable in the sense of Newton and Wigner.
Moreover, one can describe the most general expression
of the position obsevable. More conveniently, one can describe
the most general expression for the operators $B({\bf y})$
defined by (1.2). Namely, one finds out (see [2]) that:
$$(B({\bf y})f)({\bf p}) = A({\bf p})A({\bf p} + {\bf y})^{-1}
\left({E({\bf p}) \over E({\bf p} + {\bf y})} \right)^{1/2}
f({\bf p} + {\bf y}).\eqno(2.6)$$

There $A: R^{3} \rightarrow C^{2s+1}$ is a Borel function
verifying for any $U \in SU(2)$ almost everywhere:
$$D^{(s)}(U)A({\bf p}) = A(\delta (U) {\bf p})
D^{(s)}(U).\eqno(2.7)$$

Here $\delta :SU(2) \rightarrow SO(3)$ is the canonical
homomorphism. Then, according to (1.3), one finds the following
expression for the position operators $Q_{k}$:
$$(Q_{k}f)({\bf p}) = i{\partial f \over \partial p_{k}}({\bf p}) -
i{p_{k} \over E({\bf p})} \left[ {1 \over 2E({\bf p})} +
{\partial A \over \partial p_{k}}({\bf p}) A({\bf p})^{-1}\right]
f({\bf p}).\eqno(2.8)$$

2.3 We formulate now the condition of relativistic covariance
for the position operator. First, we start with the
kinematic classical picture. Suppose we have two observers
$O$ and $O'$ and they are connected by a boost of velocity
$v = th(\chi)$ in the direction ${\bf e}_{3}$. If $O'$
sees the system at $t' = 0$ in the position ${\bf x}' =
(x'_{1},x'_{2},x'_{3})$ then, according to the Lorentz
rules of transformation, $O$ sees the system localized
at ${\bf x} = (x_{1},x_{2},x_{3})$ at time $t$ where:
$$t = th(\chi)x_{3},\hskip0.5 truecm ,x'_{1} = x_{1},
\hskip0.5 truecm,x'_{2} = x_{2}, \hskip0.5 truecm,
x'_{3} = {1 \over ch(\chi)}x_{3}.\eqno(2.9)$$

When we try to formulate this result in the quantum framework
we are faced with the well-known problem that there are no states
in ${\sl H}$ strictly localized in ${\bf x}$. More precisely,
the eigenvalue equation:
$$Q_{k} \psi_{{\bf x}} = x_{k} \psi_{{\bf x}} \hskip1 truecm
(k = 1,2,3).\eqno(2.10)$$
has solution of the form:
$$\psi_{{\bf x}}({\bf p}) =  E({\bf p}) A({\bf p})v
e^{-i{\bf x} \cdot {\bf p}}.\eqno(2.11)$$
which is not an element of
${\sl H}$. Here $v$ is an arbitrary vector from $C^{2s+1}$.

Nevertheless, according to the quantum point of view, we can
consider states which are localized at ${\bf x}$ in an
approximative sense. For instance, let us define the states
$\psi_{{\bf x},\alpha,v} \in {\sl H}$
for any $\alpha \in R_{+}$ and any $v \in C^{2s+1}$, by:
$$\psi_{{\bf x},\alpha,v}({\bf p}) = \sqrt{2} \left( {2\alpha \over
\pi} \right)^{3/4} E({\bf p})^{1/2} e^{-\alpha {\bf p}^{2}}
A({\bf p})v e^{-i{\bf x} \cdot {\bf p}}.\eqno(2.12)$$
for $v \in C^{2s+1}$ of norm $1$.
Then it is easy to prove that:
$$\Vert\psi_{{\bf x},\alpha,v}\Vert^{2} = 1.\eqno(2.13)$$
and:
$$<q_{k}>_{\psi_{{\bf x},\alpha,v}} = x_{k}.\eqno(2.14)$$
$$\Delta q_{k} = \alpha^{1/2}.\eqno(2.15)$$

So, in the limit $\alpha \rightarrow 0$, $\psi_{{\bf x},\alpha,v}$
becomes a state more and more localized at ${\bf x} \in R^{3}$.

2.4 Now, let $\psi_{{\bf x}',\alpha,v}$ be a state ``aproximatively''
localized at ${\bf x}'$. We consider that this measurement is done
by the observer $O'$ at time $t' = 0$. (See Section 2.3 for
notations). According to the physical interpretation of the
representation $W^{m,+,s}$, from the point of view of the
observer $O$, the state of the system at $t = 0$ is
$W^{m,+,s}_{A_{B}(-\chi,{\bf e}_{3}),0} \psi_{{\bf x}',\alpha,v}$
where $A_{B}(\chi,{\bf n}) \in SL(2,C)$
corresponds to the boost in the direction
${\bf n}$ of velocity $v = th(\chi)$. The discussion in
Section 2.3 leads us to require that the observer
$O$ sees the particle localized around ${\bf x}$ at time
$t = th(\chi)x_{3}$. But the state of the system at time
$t = th(\chi)x_{3}$ is from the point of view of $O$:
$$\psi_{{\bf x},\alpha,v,,\chi} \equiv W^{m,+,s}_{1,th(\chi)x_{3}
e_{0}} W^{m,+,s}_{A_{B}(-\chi,{\bf e}_{3}),0} \psi_{{\bf x}',
\alpha,v}.\eqno(2.16)$$

So, our condition of manifest covariance is that $\psi_{{\bf x},\alpha,
v,\chi}$ describes a state localized around ${\bf x}$, where the connection
between ${\bf x}$ and ${\bf x}'$ is given by (2.9).
Namely, we must have:
$$<q_{k}>_{\psi_{{\bf x},\alpha,v,\chi}} = x_{k}.\eqno(2.17)$$
and:
$$lim_{\alpha \rightarrow 0} \Delta q_{k} = 0.\eqno(2.18)$$
for $k = 1,2,3$ and any $v \in C^{2s+1}$.

Let us note that if (2.17),(2.18) are true, then because of the
rotation covariance, a similar statement will be true for
boosts in any direction.

Now, it is not hard to prove that (2.17) is true for $k =1,2$
and is also true for $k = 3$ {\it iff} $A = const.$ almost
everywhere. So, we can take $A = 1$. In this case it follows easily
that $\Delta q_{k}$ behaves as $\alpha^{1/2}$ for small
$\alpha$, so (2.18) follows.

In conclusion, our main result is that the manifest covariance
condition formulated above is compatible with an {\it unique}
expression for the position operator, namely:
$$(Q_{k}f)({\bf p}) = i{\partial f \over \partial p_{k}}
({\bf p}) - i{p_{k} \over 2E({\bf p})} f({\bf p}).\eqno(2.19)$$

2.5 To study the validity of (1.4) it is more easy to work in the
vector bundle representation for $W^{m,+,s}$ (see [2], pg.365).

Then the operators $Q_{k}$ are given by the formula (VIII.230) of [2].
One can easily compute the infinitesimal generators $K_{l}$
($l = 1,2,3$) in this representation. Then, inserting apropriately
the factors $\hbar$ and $c$, one can show that we have:
$$[K_{i},Q_{j}] = 1/2 \left(Q_{i}[H,Q_{j}] +
[H,Q_{j}]Q_{i}\right) + o(\hbar c^{-2}).\eqno(2.20)$$
where $o(\hbar c^{-2})$ is an expression containing only
spin terms. This is the quantum version ot the ``manifest
covariance'' condition of Currie, Jordan and Sudarshan [12].
\vskip1 truecm
{\bf 3. The Position Operator for Non-Zero Mass Galilei System}

3.1 In the notations of [2], the system of mass $m$ and spin
$s$ corresponds to the representation $V^{\tau,s}$ of the
covering group of the proper orthochronous Galilei group
$G^{\uparrow}_{+}$. They can be realized in the Hilbert space
${\sl H} = L^{2}(R^{3},C^{2s+1},d{\bf p})$, according
to the following formula:
$$(V^{\tau,s}_{U,{\bf u},\eta,{\bf a}}f)({\bf p})=
exp \left(i{\bf a}\cdot {\bf p} +
{i \over 2} \tau {\bf a}\cdot {\bf u} -
i\eta {{\bf p}^{2} \over 2\tau}\right) D^{(s)}(U)
f(\delta(U)^{-1} ({\bf p} + \tau {\bf u})).\eqno(3.1)$$

Here $U \in SU(2),{\bf u} \in R^{3}$ is the velocity,
${\bf a}$ is the space translation, and $\eta \in R$ is
the time translation.

As in Section 2, we can describe explicitely the most general
position observable. Namely, the operators $B({\bf x})$
must have the following generic form:
$$(B({\bf x})f({\bf x}) = A({\bf p})A({\bf p}+{\bf x})^{-1}
f({\bf p}+{\bf x}).\eqno(3.2)$$
where $A:R^{3} \rightarrow C^{2s+1}$ is a Borel function
verifying (2.7). It follows that the position operators are:
$$(Q_{k}f)({\bf p}) = i{\partial f \over \partial p_{k}}
({\bf p}) - i{\partial A \over \partial p_{k}}({\bf p})
A({\bf p})^{-1} f({\bf p}).\eqno(3.3)$$

3.2 The condition of manifest covariance with respect to
Galilei boosts is simpler than in the Poincar\'e case.
Let us suppose that the two observers $O$ and $O'$ are
connected by a boost of velocity ${\bf u}$.
If $O$ sees the system at $t = 0$ in the position
${\bf x}$, then according to the formula for the Galilei boosts, the observer
$O'$ sees the system at $t' = 0$ in the position
${\bf x}' = {\bf x} - t{\bf u} = {\bf x}$

As in the Section 2, the solution of the eigenvalue equation:
$$Q_{k} \psi_{{\bf x}} = x_{k} \psi_{{\bf x}}.\eqno(3.4)$$
($k = 1,2,3$) is of the following form:
$$\psi_{{\bf x},v}({\bf p}) = e^{-i{\bf x}\cdot{\bf p}}A({\bf p})v.\eqno(3.5)$$
for any $v \in C^{2s+1}$, so is not an element of the Hilbert space
${\sl H}$. Nevertheles, we can consider the elements
$$\psi_{{\bf x},\alpha,v}({\bf p}) =
\left( {\pi \over 2\alpha}\right)^{3/4}
e^{-i{\bf p}\cdot{\bf x}}e^{-\alpha {\bf p}^{2}}A({\bf p})v.\eqno(3.6)$$
for $v \in C^{2s+1}$ of norm $1$ and we have:
$$<q_{k}>_{\psi_{{\bf x},\alpha,v}} = x_{k}.\eqno(3.7)$$
and:
$$\Delta q_{k} = \alpha^{1/2}.\eqno(3.8)$$
for $k = 1,2,3$.

3.3 Now, let $\psi_{{\bf x},\alpha,v}$ the state of the system from
the point of view of the observer $O'$. As in Section 2
the state of the system from the point of view of the
observer $O$ is:
$$\psi_{{\bf x},\alpha,{\bf u} ,v} \equiv V^{\tau,s}_{1,-{\bf u},
0,{\bf 0}} \psi_{{\bf x},\alpha,v}.\eqno(3.9)$$

According to the discussion in Section 3.2, our
condition of manifest covariance is that $\psi_{{\bf x},\alpha,{\bf u},v}$
describes a system localized around ${\bf x}$ at $t = 0$ i.e.:
$$<q_{k}>_{\psi_{{\bf x},\alpha,{\bf u},v}} = x_{k}.\eqno(3.10)$$
and:
$$lim_{\alpha \rightarrow o} \Delta q_{k} = 0.\eqno(3.11)$$
for $k = 1,2,3$.

It is elementary to prove that (3.10) is equivalent to
$A = const.$ so, as in Section 2, we can take $A = 1$.
So, in this case also, the manifest covariance condition
gives an {\it unique} position operator, namely:
$$(Q_{k}f)({\bf p}) = i{\partial f \over \partial p_{k}}
({\bf p}).\eqno(3.12)$$
It is easy to see that in this case we have:
$$[K_{i},Q_{j}] = 0.\eqno(3.13)$$

Again we note that (3.13) is the quantum counterpart of a similar classical
relation [12]. In this case there are no quantum corrections.
This can be explained by the presence of $c^{-2}$ in
the expression of the relativistic quantum correction
in (2.20).
\vskip 1truecm
{\bf 4. Conclusions}

The relation (2.20) shows that the naive quantization rules
(1.3) and (1.4) do not work always. This explains in part the
difficulties of all quantization schemes used in the literature.

It is plausible that the same unicity result hold also for the
photon if the localizability of this system is
described in the sense of Wightman framework [14].

It is interesting to try to generalize along these
lines the well known results of the ``non-interaction''
type theorems [11],15].
\vskip 1truecm
{\bf References}

\item{1.}
A. S. Wightman, ``On the Localizability of Quantum Mechanical
Systems'', Rev. Mod. Phys. 34 (1962) 845-872
\item{2.}
V. S. Varadarajan, ``Geometry of Quantum Theory'' (second edition),
Springer, 1985
\item{3.}
T. D. Newton, E. P. Wigner, ``Localized States for Elementary Systems'',
Rev. Mod. Phys. 21 (1949) 400-406
\item{4.}
A. J. Kalnay, ``The Localization Problem'' in ``Studies in the
Foundation, Methodology and Phylosophy of Sciences'', vol.4,
M. Bunge (ed.), Springer, 1971
\item{5.} J. M. Jauch, C. Piron, ``Generalized Localizability'',
Helv. Phys. Acta 40 (1967) 559-570
\item{6.} W. O. Amrein, ``Localizability for Particles of Mass Zero'',
Helv. Phys. Acta 42 (1969) 149-190
\item{7.} E. Angelopoluos, F. Bayen, M. Flato, ``On the Localizability
of Massless Particles'', Phys. Scripta 9 (1974) 173-183
\item{8.} K.Krauss, ``Position Observable for the Photon'' in
``Uncertainty Principles Principles and the Foundation of Quantum
Mechanics'' W. C. Price, S. S. Chissich (eds.) Wiley, 1977
\item{9.} A. J. Jadzyk, B. Jancewicz, ``Maximal Localizability of the
Photon'', Bull. Acad. Pol. Sci. Ser. Math. 21 (1973) 477-484
\item{10.} H. Bacry, ``Localizability and Space'', Lecture Notes
in Physics 308, Springer, 1988
\item{11.} D. G. Currie, T. F. Jordan, E. C. G. Sudarshan,
``Relativistic Invariance and Hamiltonian Theory of Interacting
Particles'', Rev. Mod. Phys. 35 (1963) 350-375
\item{12.} T. F. Jordan, N. Mukunda, ``Lorentz-Covariant Position
Operators for Spinning Particles'', Phys. Rev. 132 (1963) 1842-1848
\item{13.} J. Bertrand, ``A New Position Operator for the Photon'',
Il Nouvo Cimento 15A (1973) 281-292
\item{14.} D. R. Grigore, ``Localizability of Zero Mass Particles'',
Helv. Phys. Acta 62 (1989) 969-979
\item{15.} H. Lewtwyler, ``A No-Interaction Theorem in Classical
Relativistic Hamiltonian Particle Mechanics'', Il Nuovo Cimento
37 (1965) 556-567
\bye